%% file: main.tex
\documentclass[journal,a4paper]{IEEEtran}
\IEEEoverridecommandlockouts
\usepackage{cite}
\usepackage{amsmath,amssymb,amsfonts}
\usepackage{nicefrac}
\usepackage{algorithmic}
\usepackage{graphicx}
\ifCLASSOPTIONcompsoc
    \usepackage[caption=false, font=normalsize, labelfont=sf, textfont=sf]{subfig}
\else
\usepackage[caption=false, font=footnotesize]{subfig}
\fi
\usepackage[hidelinks]{hyperref}
\usepackage{orcidlink}
\usepackage{textcomp}
\usepackage[shortcuts]{extdash}
\usepackage{changes}
\usepackage{xcolor}
\usepackage{tikz}
\usepackage{tabularx}
\usepackage{multirow}
\usepackage{nicematrix}
\usepackage{pgfplots}
\pgfplotsset{compat=newest}
\usetikzlibrary{arrows,shapes,graphs,graphs.standard,quotes,arrows.meta,decorations.markings,positioning}
\usetikzlibrary{fit,positioning,hobby,fadings,through,shadings}
\usetikzlibrary{decorations.text}
\usetikzlibrary{decorations.pathreplacing}
\usetikzlibrary{patterns,dsp,chains,matrix,calc,backgrounds,circuits.logic.US,mux,fit}

\input{figs/register.tex}%
\definecolor{rptugrayblue} {HTML}{507289}
\definecolor{rptugraygreen}{HTML}{77b6ba}
\definecolor{rptudblue}    {HTML}{042c58}
\definecolor{rptulblue}    {HTML}{6ab2e7}
\definecolor{rptudgreen}   {HTML}{006b6b}
\definecolor{rptulgreen}   {HTML}{26d07c}
\definecolor{rptulila}     {HTML}{4c3575}
\definecolor{rptupink}     {HTML}{d13896}
\definecolor{rptured}      {HTML}{e31b4c}
\definecolor{rptuorange}   {HTML}{ffa252}

\DeclareMathOperator\sign{sign}

\DeclareMathOperator*{\argmin}{arg\,min}
\usepackage{mathtools}
\usepackage{booktabs}
\usepackage{multirow}
\usepackage[per-mode=symbol]{siunitx}
\usepackage{float}

\usepackage[acronym]{glossaries}
\loadglsentries{acronyms.tex}

\DeclareSIUnit[]{\bit}{bit}

\newcommand\mmsq[1]{\SI{#1}{\milli\meter\squared}}

\newcommand\nm[1]{\SI{#1}{\nano\meter}}
\newcommand\pc[1]{\SI{#1}{\percent}}
\newcommand{\ie}{i.\,e.,\ }

\newcommand{\til}{\raisebox{0.5ex}{\texttildelow}}
\newcommand{\IB}{\gls{IB}\ }

\newcommand{\FER}{\gls{FER}\ }
\newcommand{\FP}{\gls{FP}\ }
\newcommand{\LLR}{\gls{LLR}\ }
\newcommand{\LLRs}{\glspl{LLR}\ }
\newcommand{\LUT}{\gls{LUT}\ }
\newcommand{\LUTs}{\glspl{LUT}\ }
\newcommand{\PFT}{\gls{PFT}\ }

\newcommand{\RCQ}{\gls{RCQ}\ }
\newcommand{\REP}{\gls{REP}\ }

\newcommand{\SC}{\gls{SC}\ }

\newcommand{\SPC}{\gls{SPC}\ }
\newcommand{\SM}{\gls{SM}\ }
\newcommand{\TC}{\gls{TC}\ }

\newcolumntype{C}{>{\centering\arraybackslash}X} 
\definecolor{revcolor}{RGB}{0 0 0}

\begin{document}

\title{
High Throughput Polar Code Decoders with Information Bottleneck Quantization
}

\author{
    Claus~Kestel \orcidlink{0000-0003-3915-9097} ,~\IEEEmembership{Graduate Student Member,~IEEE,}
    Lucas~Johannsen \orcidlink{0000-0003-1858-8466},~\IEEEmembership{Graduate Student Member,~IEEE,} and
    Norbert~Wehn \orcidlink{0000-0002-9010-086X},~\IEEEmembership{Senior Member,~IEEE}
  
    \thanks{
        The authors are with the
        Microelectronic Systems Design Research Group,
        RPTU Kaiserslautern-Landau, 67663 Kaiserslautern, Germany
        (e-mail: kestel@rptu.de; lucas.johannsen@rptu.de;
        norbert.wehn@rptu.de)
    }
    \thanks{
        The authors acknowledge the financial support by the Federal Ministry of
        Education and Research of Germany in the project “Open6GHub” (grant
        number: 16KISK004).
    }
}

\maketitle

\begin{abstract}
    \input{0_abstract}
\end{abstract}

\begin{IEEEkeywords}
Forward Error Correction, Polar Code, Information Bottleneck, ASIC, 12\,nm,
Implementation
\end{IEEEkeywords}

\input{1_intro.tex}
\input{2_background}
\input{3_ib_decoder}
\input{4_results}
\input{5_conclusion}

\newpage

\bibliographystyle{IEEEtran}
\bibliography{main}
\end{document}

%% file: acronyms.tex
\newacronym{AWGN}{AWGN}{Additive White Gaussian Noise}
\newacronym{ASIC}{ASIC}{Application-Specific Integrated Circuit}
\newacronym{BER}{BER}{Bit Error Rate}
\newacronym{BP}{BP}{Belief Propagation}
\newacronym{BPSK}{BPSK}{Binary Phase Shift Keying}
\newacronym{BSMC}{BSMC}{Binary Symmetric Memoryless Channel}
\newacronym{CRC}{CRC}{Cyclic Redundancy Check}
\newacronym{FD-SOI}{FD-SOI}{Fully Depleted Silicon on Insulator}
\newacronym{FEC}{FEC}{Forward Error Correction}
\newacronym{FER}{FER}{Frame Error Rate}
\newacronym{FP}{FP}{fixed-point}
\newacronym{FSSC}{Fast-SSC}{Fast Simplified Successive Cancellation}
\newacronym{GA}{GA}{Gaussian Approximation}
\newacronym{GSM}{GSM}{Global System for Mobile Communications}
\newacronym{IP}{IP}{Intellectual Property}
\newacronym{IB}{IB}{Information Bottleneck}
\newacronym{IBM}{IBM}{Information Bottleneck Method}
\newacronym{LDPC}{LDPC}{Low Density Parity Check}
\newacronym{LLR}{LLR}{Log-Likelihood Ratio}
\newacronym{LUT}{LUT}{lookup table}
\newacronym{MSB}{MSB}{most significant bit}
\newacronym{MAP}{MAP}{Maximum a Posteriori}
\newacronym{ML}{ML}{Maximum Likelihood}
\newacronym{MS}{MS}{Min-Sum}
\newacronym{PFT}{PFT}{Polar Factor Tree}
\newacronym{PVT}{PVT}{Process, Voltage and Temperature}
\newacronym{QC}{QC}{quasi-cyclic}
\newacronym{RM}{RM}{Reed-Muller}
\newacronym{SCAN}{SCAN}{Soft Cancellation}
\newacronym{SCL}{SCL}{Successive Cancellation List}
\newacronym{SC}{SC}{Successive Cancellation}
\newacronym{SM}{SM}{Sign-Magnitude}
\newacronym{SNR}{SNR}{Signal-to-Noise-Ratio}
\newacronym{SOA}{SOA}{State-of-the-Art}
\newacronym{SRAM}{SRAM}{Static Random Access Memory}
\newacronym{SSC}{SSC}{Simplified SC}
\newacronym{SSCL}{SSCL}{Simplified SCL}
\newacronym{SoC}{SoC}{System-on-Chip}
\newacronym{TC}{TC}{Two's-Complement}
\newacronym{TDP}{TDP}{Thermal design power}
\newacronym{UMTS}{UMTS}{Universal Mobile Telecommunications System}
\newacronym{URLLC}{URLLC}{Ultra-Reliable Low Latency Communications}
\newacronym{PM}{PM}{Path Metric}
\newacronym{R0}{R0}{Rate-0}
\newacronym{RCQ}{RCQ}{Reconstruction Computation Quantization}
\newacronym{REP}{REP}{Repetition}
\newacronym{SPC}{SPC}{Single Parity-Check}
\newacronym{HD}{HD}{Hard Decision}
\newacronym{BLTA}{BLTA}{Block Lower-Triangular Affine}
\newacronym{EC}{EC}{Equivalence Class}

%% file: 0_abstract.tex
In digital baseband processing, the forward error correction (FEC) unit belongs
to the most demanding components in terms of computational complexity and power
consumption. Hence, efficient implementation of FEC decoders is crucial for next
generation mobile broadband standards and an ongoing research topic.
Quantization has a significant impact on the decoder area, power consumption and
throughput. Thus, lower bit-widths are preferred for efficient implementations
but degrade the error-correction capability. To address this issue, a
non-uniform quantization based on the Information Bottleneck (IB) method, was
proposed that enables a low bit width while maintaining the essential
information. Many investigations on the use of IB method for Low-density
parity-check code (LDPC) decoders exist and have shown its advantages from an
implementation perspective. However, for polar code decoder implementations,
there exists only one publication that is not based on the state-of-the-art
Fast-SSC decoding algorithm, and only synthesis implementation results without
energy estimation are shown. In contrast, our paper presents several optimized
Fast Simplified Successive-Cancellation (Fast-SSC) polar code decoder
implementations using IB-based quantization with placement\&routing results in
an advanced 12\,nm FinFET technology. Gains of up to 16\,\% in area and 13\,\%
in energy efficiency are achieved with IB-based quantization at a Frame Error
Rate (FER) of $10^{-7}$ and a Polar Code of $N=1024, R=0.5$ compared to
state-of-the-art decoders.

%% file: 1_intro.tex
\section{Introduction}

Polar codes are a relatively new class of \gls{FEC} codes, first described by
Erdal Ar\i kan in 2009 \cite{ari_2009}. These codes are part of the 5G
standard.
They offer low-complexity encoding and decoding algorithms, which is especially
important for high-throughput and low-latency applications in upcoming
standards such as 6G \cite{sadmeh2020}. The most commonly used decoding
algorithms for polar codes, \gls{SC} and \gls{SCL}, can be efficiently
pipelined to achieve very high throughput and low latency \cite{EPICTbps,
pimrc2020}.

Quantization has a significant impact on implementation costs. Coarse
quantization improves implementation efficiency in terms of area, power and
throughput but may decrease the error-correction performance. Finding a good
trade-off is therefore essential for efficient hardware implementations.

One promising technique to maintain the message information but enable
a reduction of the bit width is the \IB method \cite{tisper_1999}. Here, an
information compression is achieved by maximizing the mutual information between
an observed and a compressed random variable for a given bit width. This yields
a non-uniform quantization.

While \gls{IB}-based quantization for \gls{LDPC} decoder implementations is well
investigated \cite{kuryam_2008, LB18, ldpc2022}, the efficiency of the \IB
method for polar code decoder is quite unexplored. It is an open research
question whether \IB based quantization in polar decoders can yield more
efficient implementations compared to standard quantization methods.
It was shown in \cite{ldpc2022, giard2023}, that a pure \gls{LUT}-based
application of the \IB method yields very large \glspl{LUT}, making this
approach unfeasible. The \gls{RCQ} scheme \cite{wanwes_2020} also uses
\glspl{LUT} to reconstruct and, after computation, quantize back to smaller bit
width. The resulting \LUTs have much lower complexity. Hence, the \gls{RCQ}
scheme is the most promising approach from an implementation perspective.

To the best of our knowledge, only one publication investigates the efficiency
of \gls{IB}-based polar decoder implementation \cite{giard2023}. However, the
investigations in \cite{giard2023} do not consider the state-of-the-art \gls{FSSC}
decoding algorithm and, even more importantly, provides only synthesis results
without any power data, which is one of the most important implementation
metrics.

This work therefore makes the following new contributions:

\begin{itemize}
    \item We present the first \gls{FSSC} polar decoder architecture using 
        \gls{IB}-based quantization with optimized \LUTs to improve area and
        energy efficiency.
    \item We compare the non-uniform, \gls{IB}-based quantization scheme with
        uniformly quantized \FP representations in terms of error correction
        performance and implementation efficiency for code lengths of
        \SI{128}{\bit} and \SI{1024}{\bit}.
    \item We analyse the impact of the \gls{IB}-based quantization on area 
        and power consumption with 7 different designs in an advanced \nm{12}
        FinFET technology.
\end{itemize}

The remainder of this paper is structured as follows: We provide the required
background of polar codes, their decoding algorithms and the \IB method in
Sec.~\ref{sec:background}. \gls{IB}-based Fast-SSC decoding and our decoder
architecture is described in Sec.~\ref{sec:architecture}. Sec.~\ref{sec:results}
presents a detailed comparison with uniformly quantized \FP decoders in terms of
error correction performance and implementation costs and
Sec.~\ref{sec:conclusion} concludes this paper.

%% file: 2_background.tex
\section{Background}
\label{sec:background}

\subsection{Polar Codes}

Polar codes $\mathcal{P}(N,K)$ are linear block codes with code length $N=2^n$,
that encode $K$ information bits. Channel polarization derives $N$ virtual
channels where $K$ reliable channels (information set $\mathcal{I}$) are used
to transmit the information. The $N-K$ remaining (unreliable) channels are set
to zero and called frozen bits (frozen set $\mathcal{F}$). The encoding
includes a bit-reversal permutation \cite{ari_2009}.

\subsection{Successive Cancellation Decoding}

\SC decoding can be described as depth-first tree traversal of the \gls{PFT}
\cite{alaksc_2011}. The \PFT has ${\log(N)+1}$~stages~$s$ and $N$ leaf nodes at
stage $s=0$, representing the frozen and information bits. Each node~$v$
receives a \gls{LLR} vector $\boldsymbol\alpha^v$ of size~$N_s$ to first
calculate the $N_s/2$ elements of the left-child message $\boldsymbol\alpha^l$
by the hardware-efficient min-sum formulation of the $f$-function
\begin{equation}
\begin{aligned}
    \alpha^l_i &= f(\alpha^v_{2i}, \alpha^v_{2i+1})\\
    &= \sign\left(\alpha^v_{2i}\right) \sign\left(\alpha^v_{2i+1}\right)
    \min\left(\left|\alpha^v_{2i}\right|, \left|\alpha^v_{2i+1}\right| \right).
\end{aligned}\label{eq:f_fun}
\end{equation}
With the bit vector $\boldsymbol\beta^l$ received from the left child, the
$N_s/2$ elements of $\boldsymbol\alpha^r$ are calculated using the $g$-function
\begin{equation}
    \alpha^r_i = g\left(\alpha^v_{2i},\alpha^v_{2i+1}, \beta^l_i\right)
               = \left(1-2\beta^l_i\right)\cdot\alpha^v_{2i}+\alpha^v_{2i+1},
\label{eq:g_fun}
\end{equation}
and sent to the child on the right. With the results of both children,
the $h$-function calculates the partial-sum $\boldsymbol\beta^v$ with $\oplus$,
the binary XOR-operator, as
\begin{equation}
    \left(\beta^v_{2i}, \beta^v_{2i+1}\right)
    = h\left(\beta^l_i, \beta^r_i\right)
    = \left(\beta^l_{i} \oplus \beta^r_{i}, \beta^r_{i}\right),
    \label{eq:h_fun}
\end{equation}
that is sent to the parent node. In leaf nodes, bit decisions are made.
Frozen bits are $0$ per definition, and information bit nodes return the
\gls{HD} on the \gls{LLR}:
\begin{equation}
    \beta^v = \sign \left( \alpha^v \right) \triangleq
    \begin{cases}
        0 &\text{if } \alpha^v\ge 0\\
        1 &\text{otherwise.}
    \end{cases}
\end{equation}
The decoder outputs the partial-sum $\boldsymbol\beta^{0}$ of the root node.

\subsection{Fast-SSC Decoding}
\label{sec:fastssc}

Pruning the \PFT reduces the number of operations required to decode one
code word \cite{alaksc_2011}. Subtrees containing only frozen bits do not have
to be traversed, because their decoding result is known to be an all-zero
vector in advance. Such \mbox{Rate-0} nodes are always left children and are merged
into their parent nodes. Here, the $g$- and $h$-functions are executed with the
known all-zero $\boldsymbol\beta^l$, denoted by $g_0$ and $h_0$, respectively.
Similarly, subtrees without any frozen bits can be decoded directly by the
\gls{HD}, because no parity information is contained. Merging these Rate-1
nodes results in the $h_1$-function, which directly calculates
$\boldsymbol\beta^r$ using
\begin{equation}
    \beta^r_i=\sign \left(
        g\left(\alpha^v_{2i},\alpha^v_{2i+1}, \beta^l_i\right)
    \right),
    \label{eq:h1-function_b_r}
\end{equation}
and combines it with $\boldsymbol\beta^l$ to observe $\boldsymbol\beta^v$.

\gls{FSSC} decoding \cite{sargia_2014} applies further optimizations: If a
subtree contains only one information bit, it is considered a \REP code and
replaced by a specialized \REP node. All its bits are decoded by summing up the
vector $\alpha^v$ of received \gls{LLR} values and extracting the sign bit of
the sum:
\begin{equation}
    \beta^{v,\text{REP}}_i = \sign \left( \sum_{j=0}^{N_s-1} \alpha^v_j \right).
    \label{eq:rep}
\end{equation}

In subtrees containing only one frozen bit, this bit always acts as parity bit.
Thus, the partial sum of this subtree represents a \SPC code. A specialized
\SPC node performs \gls{ML} decoding by calculating the parity $\gamma^v \in
\{0,1\}$ of the input:
\begin{equation}
    \gamma^v = \bigoplus_{i=0}^{N_s-1}\sign \left( \alpha^v_i \right),
    \label{eq:parity_spc}
\end{equation}
finding the least reliable bit
\begin{equation}
    i_{\min} = \argmin_{i\in[0,N_s)}\left|\alpha^v_i\right|
    \label{eq:j_min_spc}
\end{equation}
and setting $\boldsymbol\beta^v$ to satisfy the single parity constraint:
\begin{equation}
    \beta^{v,\text{SPC}}_i = \begin{cases}
        \sign\alpha^v_i\oplus\gamma^v &\text{if } i = i_{\min}\\
        \sign\alpha^v_i               &\text{otherwise.}
              \end{cases}
    \label{eq:spc}
\end{equation}

\subsection{Information Bottleneck Method}

The \IB method is a mathematical framework used for clustering in information
theory and machine learning \cite{tisper_1999}. In the \IB
setup, the target is to preserve the shared mutual information $I(X;Y)$ between
an observed random variable $Y$ and the relevant random variable $X$ while
compressing $Y$ to $T$, \ie maximizing $I(T;X)$. Different \IB algorithms exist
\cite{slo_2002} to provide the compression mapping $p(t|y)$ derived only from
the joint distribution $p(x;y)$ and the cardinality of the compressed event
space ($|\mathcal T|$), with
$x\in\mathcal{X}=\{0,1\},y\in\mathcal{Y},t\in\mathcal{T}$ being realizations of
the random variables $X,Y,T$, respectively and $|\mathcal T| \ll |\mathcal Y|$.
A collection of \IB algorithms is provided by \cite{stalew_2018} and is used
in this work. This compression is applied to the output of an \gls{AWGN}
channel to obtain a coarse, non-uniform quantization.

In the case of our \IB decoder, $y \in\mathcal{Y}$ are the received \LLRs
that are quantized with a high bit width, e.g. \SI{10}{bit}, which equals
$|\mathcal Y| = 2^{10} = 1024 $ bins. The \IB algorithm then iteratively tries
to find pairs of bins to combine into one bin with the least loss of mutual information
$I(X;Y)$. In that way $|\mathcal Y|$ is reduced to $|\mathcal T| = 16$
(\SI{4}{bit}) and a mapping from $y$ to $t$ is derived.

%% file: 3_ib_decoder.tex
\section{IB Decoder}
\label{sec:architecture}

\subsection{Numerical Representations and LUT Generation}
\label{subsec:numerical_representation}

In this paper, we focus on decoders with very high throughput and low latency.
These decoders are fully unrolled and pipelined and use a \SM representation of
the quantized, \FP \LLR values \cite{lehall_2019}. For high \glspl{SNR},
saturation reduces signal toggling because only the sign bit changes, which
reduces power consumption. Additionally, the comparators in the $f$-functions
(\ref{eq:f_fun}) and SPC nodes (\ref{eq:j_min_spc}) can directly operate on the
magnitude. To perform the additions in the $g$-functions~(\ref{eq:g_fun}) and
REP nodes (\ref{eq:rep}), the \SM representations are converted to \TC to
efficiently perform the calculations.

Our \IB decoder implementations exploit all these optimizations. However,
because of the non-uniform distribution characterizing the \IB
indices, mathematical computations must be replaced by \LUTs \cite{giard2023}.
However, these \LUTs can become extremely large. For example, for a $g$-function
(\ref{eq:g_fun}) with one binary and two \LLR inputs, the resulting \LUT is of
size $2\cdot|\mathcal T|^2$.
From an implementation perspective, the \LUTs are transformed into Boolean
functions. Despite the logic optimization executed by state-of-the-art synthesis
tools, the resulting logic costs can quickly outweigh the benefits of reduced
bit widths, particularly for increasing \LUT sizes \cite{ldpc2022}.

A promising approach to address this problem is the \gls{RCQ} scheme
\cite{wanwes_2020, mohsha_2023}. \gls{RCQ} combines non-uniform quantization
with the traditional node computations. Only small \LUTs are necessary that are
placed in front of the computation units to upscale the reduced non-uniform \IB
quantization to a uniform \FP quantization (\emph{Reconstruction}). Node computations are then
performed on the uniform \FP quantization (\emph{Computation}). After computation the results
are downscaled to smaller bit width (\emph{Quantization}). The \gls{RCQ} scheme results in
much smaller \LUTs because the mappings for the conversions between \IB and
\FP domains can be implemented as \LUTs of size $2^Q$ for each value where
$Q$ is the bit width. We use $Q_\text{IB} = \log_2(|\mathcal T|)$ and
$Q_\text{FP}$ to denote the bit widths in the \IB and \FP domains,
respectively. For the $g$-function example mentioned above, the number of entries in the \LUTs shrinks
from $2\cdot|\mathcal T|^2=2^{1+2\cdot Q_\text{IB}}$ to $2\cdot2^{Q_\text{IB}}
+ 2^{Q_\text{FP}}$.

\begin{table}
    \centering
    \caption{
        Hardware-aware representation of \SI{3}{\bit} IB indices
    }\label{tab:mapping}
    \input{tables/mapping.tex}
\end{table}

We use density evolution to generate samples. At least 100K samples are
monitored at each node in the decoder, \ie at each edge of the \gls{PFT}. These
samples yield the joint distributions $p(x;y)$ which are input to the \IB
algorithm \cite{slo_2002} that calculates the compression mappings $p(t|y)$ for
every edge. Then, the \textit{Symmetric information bottleneck algorithm}
\cite{LB18, stalew_2018} is applied which is optimized to preserve the symmetry
of the transmission channel (assuming an \gls{AWGN} channel and \gls{BPSK}
modulation). Exploiting this symmetry enables a bisection of the \LUT, because
it is sufficient to store only the magnitudes. Accordingly, we use a
\gls{SM}-like representation of the \IB indices as shown in the example in
Table~\ref{tab:mapping} for $|\mathcal T| = 8$.
Therefore, for both directions of conversion (\IB to \gls{FP}, and \FP to
\gls{IB}), the \LUTs directly map one magnitude to another, \ie the magnitudes
in both domains also act as indices for the \glspl{LUT}. Thus, the size of each
\LUT is $2^{Q-1}$ 
and the total
number of entries for the example of the $g$-function becomes $2\cdot2^{Q_\text{IB}-1} +
2^{Q_\text{FP}-1}$. As shown in Figure \ref{fig:g_fun} this means a reduction
from $2 * 16^2 = 512$ entries to $2*2^{4-1}$ for the upscaling \LUT and $2^{5-1}$
for the downscaling LUT, making it just 32 entries.

Furthermore, this approach eliminates the need for multiple
comparisons with thresholds per conversion as in \cite{mohsha_2023}.

\subsubsection*{Notation}

To differentiate between the numerical representations, we use $\alpha$ to
denote values in the \IB domain, $\tilde{\alpha}$ for \SM and $\ddot{\alpha}$
for \TC representation. The $j$-th bit of the binary expansion of $\alpha_i$ is
given by $\alpha_{i_{(j)}}$ and the \gls{MSB} $Q-1$ refers to the sign bit.

\subsection{IB-based Fast-SSC Decoding}

\subsubsection{$f$-Function}

With the symmetric mapping and inherent order of the \IB indices
(Table~\ref{tab:mapping}), the $f$-function (\ref{eq:f_fun}) can be directly
performed in the non-uniform \IB domain and no \LUTs are necessary, which corresponds to the
``\emph{re-MS-IB} decoder'' implementation of \cite{giard2023}. In contrast to
\cite{giard2023}, we map the \IB indices so that negative values also
correspond to negative \glspl{LLR} and, thus, do not need to invert the result
of the \emph{XOR}-ed sign bits.

\subsubsection{$g$-Function}

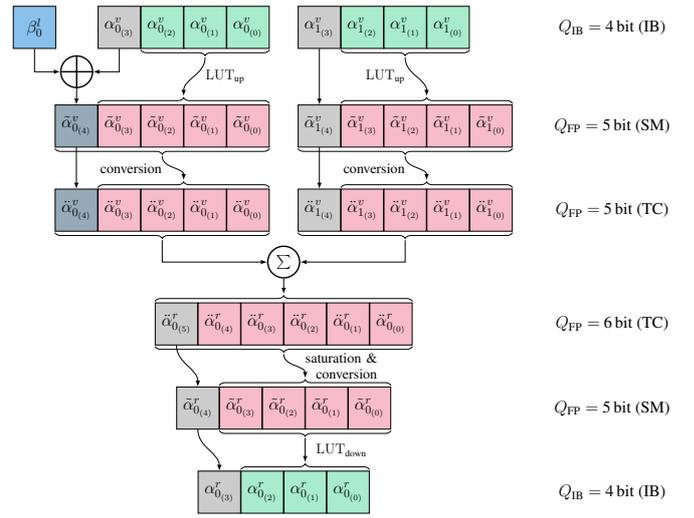
\begin{figure}
    \centering
    \resizebox{\columnwidth}{!}{\input{figs/g_fun}}
    \caption{\RCQ schematic for the $g$-function}
    \label{fig:g_fun}
\end{figure}

As in \cite{mohsha_2023}, we apply an \RCQ scheme, but based on our optimized
up- and downscale \glspl{LUT}:
\begin{equation}
    \alpha^r_i = \mathrm{LUT}^v_\text{down}\left(
        g\left(
            \mathrm{LUT}^v_\text{up}\left(\alpha^v_{2i}\right),
            \mathrm{LUT}^v_\text{up}\left(\alpha^v_{2i+1}\right),
            \beta^l_i
        \right)
    \right).
\end{equation}

The internal separation between the different number representations is
maintained for the reasons described in
Sec.~\ref{subsec:numerical_representation} and shown in Fig.~\ref{fig:g_fun}. The
\emph{Reconstruction} with $\mathrm{LUT}^v_\text{up}$ maps the magnitude of the
\IB index $\alpha^v_i$ to its \SM representation $\tilde{\alpha}^v_i$ with
preserved sign bit. $\beta^l_i$ must be considered before conversion to the \TC
representation $\ddot{\alpha}^v_i$ for the \emph{Computation} step. The result
$\ddot{\alpha}^r_i$ has a \SI{1}{\bit} larger resolution, implying a saturation
for the conversion back to \SM representation with
\SI[parse-numbers=false]{Q_\text{FP}}{\bit}. The \emph{Quantization} step is
again realized as magnitude $\mathrm{LUT}^v_\text{down}$ for the transformation
back to the \IB domain.

For the special case of merged Rate-0 nodes, \ie the $g_0$-function, the
\emph{XOR}-operation with $\beta^v_i$ is omitted.

\subsubsection{REP Nodes}

\REP nodes calculate the sum over all input values to observe the single
(repeated) information bit by \gls{HD} on the sum (\ref{eq:rep}). Thus, the
\RCQ scheme as described for the $g$-function is applied for \REP nodes. An
adder tree of $N_s-1$ adders and a depth of $s=\log_2 N_s$ operates on the \TC
representations of the $N_s$ input values. The final \gls{HD} as
\emph{Quantization} step extracts the single sign bit as bit decision of the
node, for which reason no further conversion of the sum is needed.

\subsubsection{SPC Nodes}

As in the f-function, due to the ordered mapping of the \IB indices, the
minimum search of the \SPC node (\ref{eq:j_min_spc}) can be performed directly
in the IB domain. Furthermore, the chosen mapping is also suitable for the
direct parity calculation (\ref{eq:parity_spc}) and the bit estimations
(\ref{eq:spc}), because the sign bits are preserved in the \IB domain.

\subsubsection{$h_{1}$-Function}

The $h_1$-function internally uses the $g$-function to compute $\beta^r$
(\ref{eq:h1-function_b_r}). However, in this $g$-function, the backward
conversion is not needed, because the \gls{HD} is made directly in the compute
domain with \TC representation, as already described for the \REP nodes.

\subsection{Decoder Architecture}
\label{sec:dec_arch}

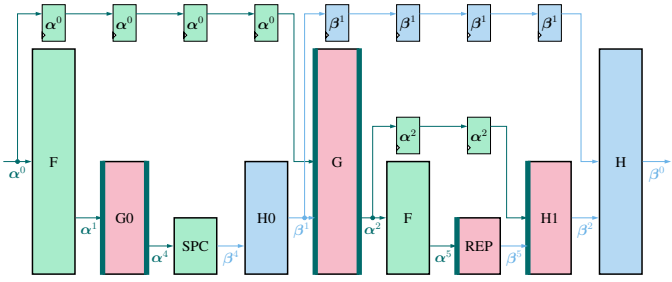
\begin{figure}
    \centering
    \resizebox{\columnwidth}{!}{\input{figs/unrolled_sc}}
    \caption{
        Unrolled and pipelined Fast-SSC decoder architecture for a
        $\mathcal{P}(16,8)$. Colors represent numerical domains: green for IB,
        red for FP, blue for binary and dark green shows the LUTs.
    }
    \label{fig:unrolled_sc}
\end{figure}

An outline of the fully unrolled and deeply pipelined Fast-\gls{SSC} decoder
architecture for a $\mathcal P(16,8)$ is shown in Figure ~\ref{fig:unrolled_sc}. We
omit the clock signals and the delay lines are represented by shift
registers. The pipeline consists of various building blocks that implements the
decoding functions described in the previous section. The IB, \gls{FP} and
binary domains are represented by the coloring of the blocks and signals. The
decoders presented in this paper are based on the framework presented in
\cite{lehall_2019} which we extended to apply the \IB method as described above.

%% file: tables/mapping.tex
\begin{tabular}{ |c|c|c|c|c|c|c|c|c| } 
 \hline
 IB Index $t$   &     0 &     1 &     2 &     3 &     4 &     5 &     6 &     7 \\ 
 \hline
 \SM IB   &    -3 &    -2 &    -1 &    -0 &     0 &     1 &     2 &     3 \\ 
 \hline
 Binary   & 1\,11 & 1\,10 & 1\,01 & 1\,00 & 0\,00 & 0\,01 & 0\,10 & 0\,11 \\ 
 \hline
\end{tabular}

%% file: figs/g_fun.tex
\colorlet{IB}{rptulgreen!40}%
\colorlet{FP}{rptured!30}%
\colorlet{bit}{rptulblue!80}%
\tikzset{
mymat/.style={
  matrix of nodes,
  anchor=center,
  font=\Large,
  minimum size=8ex,
  },%
mymats/.style={
  mymat,
  nodes={fill=#1, draw},
  column 1/.style={nodes={fill=white!80!black}}
  },%
description/.style={font=\Large},%
cross/.style={path picture={\draw[black]
    (path picture bounding box.south)--(path picture bounding box.north)
    (path picture bounding box.west) -- (path picture bounding box.east);
    }},%
crossx/.style={path picture={\draw[black]
    (path picture bounding box.south west)--(path picture bounding box.north east)
    (path picture bounding box.north west)--(path picture bounding box.south east);
    }},%
every picture/.append style={font=\large},%
}
\begin{tikzpicture}[>=latex]
\matrix[mymats=IB] at (0,0) (alpha_r_ib) {
$\alpha^r_{0_{(3)}}$ &
$\alpha^r_{0_{(2)}}$ &
$\alpha^r_{0_{(1)}}$ &
$\alpha^r_{0_{(0)}}$ \\
};
\matrix[mymats=FP, above=of alpha_r_ib] (alpha_r_sm) {
$\tilde{\alpha}^r_{0_{(4)}}$ &
$\tilde{\alpha}^r_{0_{(3)}}$ &
$\tilde{\alpha}^r_{0_{(2)}}$ &
$\tilde{\alpha}^r_{0_{(1)}}$ &
$\tilde{\alpha}^r_{0_{(0)}}$ \\
};
\matrix[mymats=FP, above=of alpha_r_sm] (alpha_r_tc) {
$\ddot{\alpha}^r_{0_{(5)}}$ &
$\ddot{\alpha}^r_{0_{(4)}}$ &
$\ddot{\alpha}^r_{0_{(3)}}$ &
$\ddot{\alpha}^r_{0_{(2)}}$ &
$\ddot{\alpha}^r_{0_{(1)}}$ &
$\ddot{\alpha}^r_{0_{(0)}}$ \\
};
\node[draw, circle,
    very thick, minimum size=6ex, above=6mm of alpha_r_tc] (plus) {$\sum$};
\coordinate[above=3mm of plus.north] (tmp);
\matrix[mymats=FP, left=3mm of tmp, anchor=south east,
  column 1/.style={nodes={fill=rptugrayblue!60!white}}
] (alpha_v_0_tc) {
$\ddot{\alpha}^v_{0_{(4)}}$ &
$\ddot{\alpha}^v_{0_{(3)}}$ &
$\ddot{\alpha}^v_{0_{(2)}}$ &
$\ddot{\alpha}^v_{0_{(1)}}$ &
$\ddot{\alpha}^v_{0_{(0)}}$ \\
};
\matrix[mymats=FP, right=3mm of tmp, anchor=south west] (alpha_v_1_tc) {
$\ddot{\alpha}^v_{1_{(4)}}$ &
$\ddot{\alpha}^v_{1_{(3)}}$ &
$\ddot{\alpha}^v_{1_{(2)}}$ &
$\ddot{\alpha}^v_{1_{(1)}}$ &
$\ddot{\alpha}^v_{1_{(0)}}$ \\
};
\matrix[mymats=FP, above=of alpha_v_0_tc,
  column 1/.style={nodes={fill=rptugrayblue!60!white}}
] (alpha_v_0_sm) {
$\tilde{\alpha}^v_{0_{(4)}}$ &
$\tilde{\alpha}^v_{0_{(3)}}$ &
$\tilde{\alpha}^v_{0_{(2)}}$ &
$\tilde{\alpha}^v_{0_{(1)}}$ &
$\tilde{\alpha}^v_{0_{(0)}}$ \\
};
\matrix[mymats=FP,above=of alpha_v_1_tc] (alpha_v_1_sm) {
$\tilde{\alpha}^v_{1_{(4)}}$ &
$\tilde{\alpha}^v_{1_{(3)}}$ &
$\tilde{\alpha}^v_{1_{(2)}}$ &
$\tilde{\alpha}^v_{1_{(1)}}$ &
$\tilde{\alpha}^v_{1_{(0)}}$ \\
};
\matrix[mymats=IB, above=22mm of alpha_v_0_sm.east, anchor=south east,
] (alpha_v_0_ib) {
$\alpha^v_{0_{(3)}}$ &
$\alpha^v_{0_{(2)}}$ &
$\alpha^v_{0_{(1)}}$ &
$\alpha^v_{0_{(0)}}$ \\
};
\matrix[mymats=IB, above=22mm of alpha_v_1_sm.west, anchor=south west] (alpha_v_1_ib) {
$\alpha^v_{1_{(3)}}$ &
$\alpha^v_{1_{(2)}}$ &
$\alpha^v_{1_{(1)}}$ &
$\alpha^v_{1_{(0)}}$ \\
};
\matrix[mymats=white, left=of alpha_v_0_ib,
  column 1/.style={nodes={fill=bit}}
] (beta_l_0) {
$\beta^l_{0}$ \\
};
\coordinate (tmp) at ($(beta_l_0)!0.5!(alpha_v_0_ib-1-1)$);
\node[draw, circle, cross, 
    very thick, minimum size=6ex] at ($(tmp)!0.45!(alpha_v_0_sm-1-1)$) (xor) {};
\node[description, right=of alpha_v_1_sm]     (sm_in) {$Q_\text{FP}=$ \SI{5}{\bit} (SM)};
\node[description] at (sm_in |- alpha_v_1_ib) (ib_in) {$Q_\text{IB}=$ \SI{4}{\bit} (IB)};
\node[description] at (sm_in |- alpha_v_1_tc) (tc_in) {$Q_\text{FP}=$ \SI{5}{\bit} (TC)};
\node[description] at (sm_in |- alpha_r_tc)   (tc_out){$Q_\text{FP}=$ \SI{6}{\bit} (TC)};
\node[description] at (sm_in |- alpha_r_sm)   (sm_out){$Q_\text{FP}=$ \SI{5}{\bit} (SM)};
\node[description] at (sm_in |- alpha_r_ib)   (ib_out){$Q_\text{IB}=$ \SI{4}{\bit} (IB)};

\begin{scope}[
    thick
]
\draw[->] (beta_l_0-1-1.south)     |- (xor.west);
\draw[->] (alpha_v_0_ib-1-1.south) |- (xor.east);
\draw[->] (xor.south) -- (alpha_v_0_sm-1-1.north);
\draw[->] (alpha_v_0_sm-1-1.south) -- (alpha_v_0_tc-1-1.north);

\draw[->] (alpha_v_1_ib-1-1.south) -- (alpha_v_1_sm-1-1.north);
\draw[->] (alpha_v_1_sm-1-1.south) -- (alpha_v_1_tc-1-1.north);

\draw[decorate,decoration={brace,raise=1pt,amplitude=5pt,mirror}, thick]
    (alpha_v_0_ib-1-2.south west) -- node[below](tmp1){} (alpha_v_0_ib-1-4.south east);
\draw[decorate,decoration={brace,raise=1pt,amplitude=5pt}, thick]
    (alpha_v_0_sm-1-2.north west) -- node[above](tmp2){} (alpha_v_0_sm-1-5.north east);
\draw[->] (tmp1) to[out=-90,in=90] node[right=2mm]{$\mathrm{LUT}_\text{up}$} (tmp2);

\draw[decorate,decoration={brace,raise=1pt,amplitude=5pt,mirror}, thick]
    (alpha_v_1_ib-1-2.south west) -- node[below](tmp1){} (alpha_v_1_ib-1-4.south east);
\draw[decorate,decoration={brace,raise=1pt,amplitude=5pt}, thick]
    (alpha_v_1_sm-1-2.north west) -- node[above](tmp2){} (alpha_v_1_sm-1-5.north east);
\draw[->] (tmp1) to[out=-90,in=90] node[left=2mm]{$\mathrm{LUT}_\text{up}$} (tmp2);

\draw[decorate,decoration={brace,raise=1pt,amplitude=5pt,mirror}, thick]
    (alpha_v_0_sm-1-1.south west) -- node[below](tmp1){} (alpha_v_0_sm-1-5.south east);
\draw[decorate,decoration={brace,raise=1pt,amplitude=5pt}, thick]
    (alpha_v_0_tc-1-2.north west) -- node[above](tmp2){} (alpha_v_0_tc-1-5.north east);
\draw[->] (tmp1) to[out=-90,in=90] node[left=2mm]{conversion} (tmp2);

\draw[decorate,decoration={brace,raise=1pt,amplitude=5pt,mirror}, thick]
    (alpha_v_1_sm-1-1.south west) -- node[below](tmp1){} (alpha_v_1_sm-1-5.south east);
\draw[decorate,decoration={brace,raise=1pt,amplitude=5pt}, thick]
    (alpha_v_1_tc-1-2.north west) -- node[above](tmp2){} (alpha_v_1_tc-1-5.north east);
\draw[->] (tmp1) to[out=-90,in=90] node[left=2mm]{conversion} (tmp2);

\draw[decorate,decoration={brace,raise=1pt,amplitude=5pt,mirror}, thick]
    (alpha_v_0_tc-1-1.south west) -- node[below](tmp0){} (alpha_v_0_tc-1-5.south east);
\draw[decorate,decoration={brace,raise=1pt,amplitude=5pt,mirror}, thick]
    (alpha_v_1_tc-1-1.south west) -- node[below](tmp1){} (alpha_v_1_tc-1-5.south east);
\draw[->] (tmp0) |- (plus.west);
\draw[->] (tmp1) |- (plus.east);

\draw[decorate,decoration={brace,raise=1pt,amplitude=5pt}, thick]
    (alpha_r_tc-1-1.north west) -- node[above](tmp2){} (alpha_r_tc-1-6.north east);
\draw[->] (plus.south) -- (tmp2);

\draw[decorate,decoration={brace,raise=1pt,amplitude=5pt, mirror}, thick]
    (alpha_r_tc-1-1.south west) -- node[below](tmp1){} (alpha_r_tc-1-6.south east);
\draw[decorate,decoration={brace,raise=1pt,amplitude=5pt}, thick]
    (alpha_r_sm-1-2.north west) -- node[above](tmp2){} (alpha_r_sm-1-5.north east);
\draw[->] (tmp1) to[out=-90,in=90] node[right=2mm, align=right]{saturation \&\\conversion} (tmp2);
\draw[->] (alpha_r_tc-1-1.south) to[out=-90,in=90] (alpha_r_sm-1-1.north);

\draw[decorate,decoration={brace,raise=1pt,amplitude=5pt, mirror}, thick]
    (alpha_r_sm-1-2.south west) -- node[below](tmp1){} (alpha_r_sm-1-5.south east);
\draw[decorate,decoration={brace,raise=1pt,amplitude=5pt}, thick]
    (alpha_r_ib-1-2.north west) -- node[above](tmp2){} (alpha_r_ib-1-4.north east);
\draw[->] (tmp1) -- node[right=2mm]{$\mathrm{LUT}_\text{down}$} (tmp2);
\draw[->] (alpha_r_sm-1-1.south) to[out=-90,in=90] (alpha_r_ib-1-1.north);

\end{scope}
\end{tikzpicture}

%% file: figs/unrolled_sc.tex
\colorlet{IB}{rptudgreen}%
\colorlet{IBreg}{rptulgreen!40}%
\colorlet{FP}{rptured!30}%
\colorlet{llr}{IB}%
\colorlet{llrreg}{IBreg}%
\colorlet{bit}{rptulblue}%
\colorlet{bitreg}{rptulblue!50}%
\colorlet{RR}{rptugrayblue!50}%
\colorlet{REP}{rptured!70}%
\colorlet{SPC}{rptulgreen!70}%
\colorlet{R0}{white}%
\colorlet{R1}{black}%
\colorlet{mI}{rptulblue}%
\colorlet{mIreg}{rptulblue!40}%
\colorlet{dF}{rptuorange!70}%
\colorlet{path}{rptulila}%
\colorlet{pathreg}{rptulila!40}%
\colorlet{text}{black}%
\tikzset{%
    stageR/.style = {draw, very thick, rectangle, fill=IBreg,%
                    minimum width = 12mm, align=center, anchor=south},%
    stageGR/.style = {stageR, fill=FP},
    stageG0R/.style = {stageR, fill=FP},
    stageHR/.style = {stageR, fill=bitreg},%
    stageH0R/.style = {stageR, fill=bitreg},%
    stageH1R/.style = {stageR, fill=FP},%
    repR/.style   = {stageGR, fill=FP},%
    spcR/.style   = {stageGR, fill=IBreg},%
    stageC/.style = {stageR, dotted},%
    rate0C/.style = {rate0R, dotted},%
    rate1C/.style = {rate1R, dotted},%
    repC/.style   = {repR,   dotted},%
    spcC/.style   = {spcR,   dotted},%
    dot/.style   = {dspnodefull,minimum size=3pt,inner sep=0pt},%
    regR/.style   = {reg, circuit symbol size=width 8 height 1},%
    regC/.style   = {reg, dotted},%
    sig/.style    = {font=\sffamily\large},
    every picture/.append style={font=\large},%
}%
\begin{tikzpicture}[circuit logic US, thick]%
    \node[stageR,   minimum height=64mm] at (0,0)                          (0F){F};
    \node[stageG0R, minimum height=32mm, right=2 of 0F.south, anchor=south](1G){G0};
    \draw[line width=1.5mm, IB] (1G.north west) -- (1G.south west);
    \draw[line width=1.5mm, IB] (1G.north east) -- (1G.south east);
    \node[spcR,     minimum height=16mm, right=2 of 1G.south, anchor=south](4R){SPC};
    \node[stageH0R, minimum height=32mm, right=2 of 4R.south, anchor=south](1H){H0};
    \node[stageGR,  minimum height=64mm, right=2 of 1H.south, anchor=south](0G){G};
    \draw[line width=1.5mm, IB] (0G.north west) -- (0G.south west);
    \draw[line width=1.5mm, IB] (0G.north east) -- (0G.south east);
    \node[stageR,   minimum height=32mm, right=2 of 0G.south, anchor=south](2F){F};
    \node[repR,     minimum height=16mm, right=2 of 2F.south, anchor=south](5S){REP};
    \draw[line width=1.5mm, IB] (5S.north west) -- (5S.south west);
    \node[stageH1R, minimum height=32mm, right=2 of 5S.south, anchor=south](2H){H1};
    \draw[line width=1.5mm, IB] (2H.north west) -- (2H.south west);
    \node[stageHR,  minimum height=64mm, right=2 of 2H.south, anchor=south](0H){H};
    
    \coordinate[left =8mm of 0F.west] (input);
    \coordinate[right=8mm of 0H.east] (output);
    \draw[dspconn, llr, thick] (input)           --node[sig,below]{$\boldsymbol\alpha^0$} (0F.west);
    \draw[dspconn, llr, thick] (0F.east|-1G.west)--node[sig,below]{$\boldsymbol\alpha^1$} (1G.west);
    \draw[dspconn, llr, thick] (1G.east|-4R.west)--node[sig,below]{$\boldsymbol\alpha^4$} (4R.west);
    \draw[dspconn, llr, thick] (0G.east|-2F.west)--node[sig,below]{$\boldsymbol\alpha^2$} (2F.west);
    \draw[dspconn, llr, thick] (2F.east|-5S.west)--node[sig,below]{$\boldsymbol\alpha^5$} (5S.west);
    
    \draw[dspconn, bit, thick] (4R.east)--node[sig,below]{$\boldsymbol\beta^4$} (4R.east-|1H.west);
    \draw[dspconn, bit, thick] (1H.east)--node[sig,below]{$\boldsymbol\beta^1$} (1H.east-|0G.west);
    \draw[dspconn, bit, thick] (5S.east)--node[sig,below]{$\boldsymbol\beta^5$} (5S.east-|2H.west);
    \draw[dspconn, bit, thick] (2H.east)--node[sig,below]{$\boldsymbol\beta^2$} (2H.east-|0H.west);
    \draw[dspconn, bit, thick] (0H.east)--node[sig,below]{$\boldsymbol\beta^0$} (output);    
    \node[regR, fill=llrreg, above=2mm of 0F]            (REGa00) {$\boldsymbol \alpha^0$};
    \node[regR, fill=llrreg] at (REGa00.east -| 1G.north)(REGa01) {$\boldsymbol \alpha^0$};
    \node[regR, fill=llrreg] at (REGa01.east -| 4R.north)(REGa02) {$\boldsymbol \alpha^0$};
    \node[regR, fill=llrreg] at (REGa02.east -| 1H.north)(REGa03) {$\boldsymbol \alpha^0$};
    
    \node[regR, fill=llrreg, above=2mm of 2F]            (REGa20) {$\boldsymbol \alpha^2$};
    \node[regR, fill=llrreg] at (REGa20.east -| 5S.north)(REGa21) {$\boldsymbol \alpha^2$};
    
    \node[regR, fill=bitreg, above=2mm of 0G]            (REGb10) {$\boldsymbol \beta^1$};
    \node[regR, fill=bitreg] at (REGb10.east -| 2F.north)(REGb11) {$\boldsymbol \beta^1$};
    \node[regR, fill=bitreg] at (REGb11.east -| 5S.north)(REGb12) {$\boldsymbol \beta^1$};
    \node[regR, fill=bitreg] at (REGb12.east -| 2H.north)(REGb13) {$\boldsymbol \beta^1$};
    
    \path[overlay] (input|-0F.west) -- (0F.west) node[midway, dot, llr](tmp){};
    \draw[dspconn, llr, thick] (tmp) |- (REGa00.D);
    \draw[dspconn, llr, thick] (REGa00.Q) -- (REGa01.D);
    \draw[dspconn, llr, thick] (REGa01.Q) -- (REGa02.D);
    \draw[dspconn, llr, thick] (REGa02.Q) -- (REGa03.D);
    \draw[dspconn, llr, thick] (REGa03.Q) -| ($(0G.west)-(6mm,0)$) |- (0G.west);
    
    \path[overlay] (0G.east|-2F.west) -- (2F.west) node[midway, dot, llr](tmp){};
    \draw[dspconn, llr, thick] (tmp) |- (REGa20.D);
    \draw[dspconn, llr, thick] (REGa20.Q) |- (REGa21.D);
    \draw[dspconn, llr, thick] (REGa21.Q) -| ($(2H.west)-(6mm,0)$) |- (2H.west);
    
    \coordinate (tmp) at (1H.east-|0G.west);
    \node[dot, bit] at ($(tmp)-(3mm,0)$) (tmp) {};
    \draw[dspconn, thick, bit] (tmp) |- (REGb10.D);
    \draw[dspconn, thick, bit] (REGb10.Q) -- (REGb11.D);
    \draw[dspconn, thick, bit] (REGb11.Q) -- (REGb12.D);
    \draw[dspconn, thick, bit] (REGb12.Q) -- (REGb13.D);
    \coordinate (tmp) at ($(0H.west)-(5mm,0)$);
    \draw[dspconn, thick, bit] (REGb13.Q) -| (tmp) |- (0H.west);
    
\end{tikzpicture}

%% file: 4_results.tex
\section{Results}
\label{sec:results}

We present 7 decoder designs for $N\!=\!128$ and $N\!=\!1024$ which are
optimized for a target frequency of \SI{500}{\mega\hertz} and
\SI{750}{\mega\hertz}, respectively. Throughput is considered as coded
throughput. The designs were synthesized with \textit{Design Compiler} and
placed and routed with \textit{IC-Compiler}, both from \textit{Synopsys} in a
\nm{12} FinFET technology with a super-low~$V_t$ transistor library from
\textit{Global Foundries} under worst case \gls{PVT} conditions
(\SI{125}{\degreeCelsius}, \SI{0.72}{\volt}) for timing and nominal case
\gls{PVT} (\SI{25}{\degreeCelsius}, \SI{0.8}{\volt}) for power. Error-correction
performance was simulated for an \gls{AWGN} channel and \gls{BPSK} modulation
with a minimum of 100 erroneous code words.

\subsection{Decoder for the Polar(128,64) Code}

\begin{figure}[h]
	\includegraphics[width=0.95\linewidth]{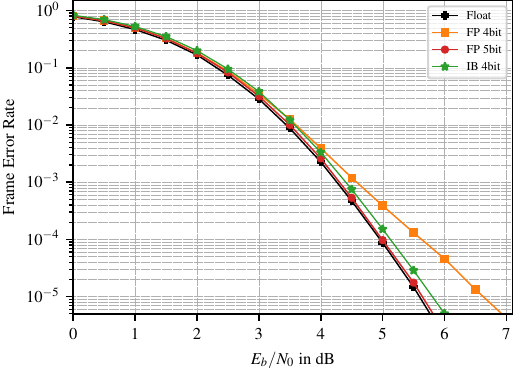}
	\centering
    \caption{\FER of the $\mathcal{P}(128,64)$, Float vs. FP vs. IB Quantization}
    \label{fig:plot_128}
\end{figure}

The \FER of the $\mathcal{P}(128,64)$ is shown in Figure~\ref{fig:plot_128}. The
\IB decoder with \SI{4}{\bit} and the \FP decoder with \SI{5}{\bit} show
similar error-correction performance, whereas the error-correction performance
of the \SI{4}{\bit} \FP decoder starts to degrade at an \FER of $10^{-3}$.

\begin{table*}[t]
    \centering
    \caption{
        Implementation results for $\mathcal{P}(128,64)$ decoders}
    \label{tab:128}
    \input{tables/results_128.tex}
      \noindent{\footnotesize{\textsuperscript{*} Synthesis only, scaled from \nm{28} to \nm{12} (numbers in brackets) with equations from  \cite{STILLMAKER201774} and maximum frequency limited to \SI{1000}{\mega\hertz}}. For \nm{28} and \nm{12} the scaling factors of \nm{32} and \nm{14} were used, since they belong to the same technology generation and give the best approximation.}
\end{table*}

Table~\ref{tab:128} presents the corresponding implementation results.
Comparing the decoders with similar \FER (\IB vs. \SI{5}{\bit} \gls{FP}), we
observe similar combinatorial area (logic), whereas the area for the memory
(registers) is reduced by \til\SI{16}{\percent}. This improves the area
efficiency by \til\SI{7}{\percent} and yields power savings and energy
efficiency improvements of \til\SI{13}{\percent}.

When comparing the \IB decoder with the \SI{4}{\bit} \FP decoder, the cost
of the \LUTs can be directly observed in the combinatorial area (0.005 mm$^2$ vs.
0.003mm$^2$) while the area for the registers is identical. This cost can be
considered as the price for the improved error correction performance of the \IB
decoder.

As already mentioned there exits only one other publication that gives
implementation results for \gls{IB}-based decoders. Since the authors of
\cite{giard2023} only provide synthesis results in an older \nm{28} technology,
a fair comparison is difficult. To enable at least some comparison, we scaled
the results of \cite{giard2023} to \nm{12} according to the equations of
\cite{STILLMAKER201774}. The scaled results are included in Table \ref{tab:128}.
We limited the maximum frequency to \SI{1000}{\mega\hertz} which is more
reasonable, since \SI{3681}{\mega\hertz} (and even \SI{1510}{\mega\hertz} in the
original publication) are not realistic for a design in \nm{28} after placement.
The reasons are: first, the power consumption and power density becomes
infeasible with this high frequency. Second, \SI{3681}{\mega\hertz} are
unfeasible for standard placement and routing in semi-custom design flows in
\nm{28} technology. Even with the scaled estimate without frequency limitation,
we observe that our optimised decoders outperform \cite{giard2023} in
throughput, latency, area and area efficiency.

\subsection{Decoder for the Polar(1024,512) Code}

\begin{figure}[h]
	\includegraphics[width=0.95\linewidth]{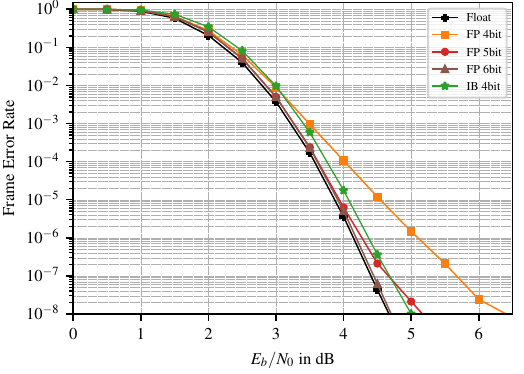}
	\centering
    \caption{
        \FER of the $\mathcal{P}(1024,512)$, Float vs. \FP vs. \IB 
    }
    \label{fig:plot_1024}
\end{figure}

The longer $\mathcal{P}(1024,512)$ polar code has a longer pipeline and is
therefore more affected by accumulating quantization errors. In contrast to the
shorter code, \SI{6}{\bit} \FP are necessary to match the floating point
precision. We compare the \SI{4}{\bit} \IB decoder to the \FP decoder with
\SI{5}{\bit}, as they show similar error-correction performance (Figure
\ref{fig:plot_1024}). Here, the \IB decoder even outperforms the \SI{5}{\bit}
\FP decoder at an \FER of $10^{-7}$.

\begin{table}
    \centering
    \caption{
        Implementation results for $\mathcal{P}(1024,512)$ decoders}
    \label{tab:1024}
    \input{tables/results_1024.tex}
    
\end{table}

\begin{figure}
    \centering
    \subfloat[FP \SI{5}{\bit}: \mmsq{0.968}]{%
        \includegraphics[height=0.4919\columnwidth] {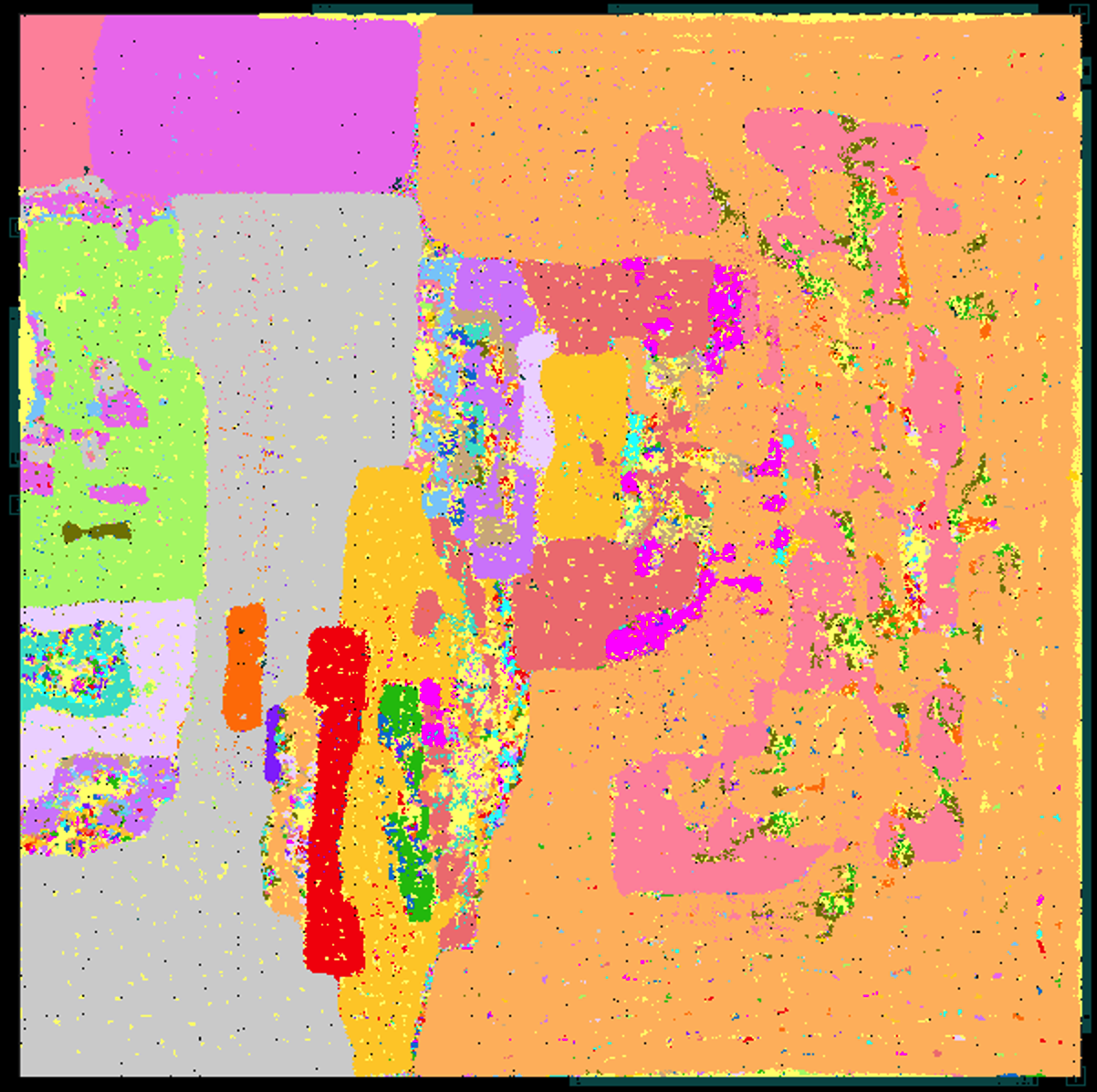}%
    }
    \hfill
    \subfloat[IB \SI{4}{\bit}: \mmsq{0.822}]{%
        \includegraphics[height=0.4533\columnwidth] {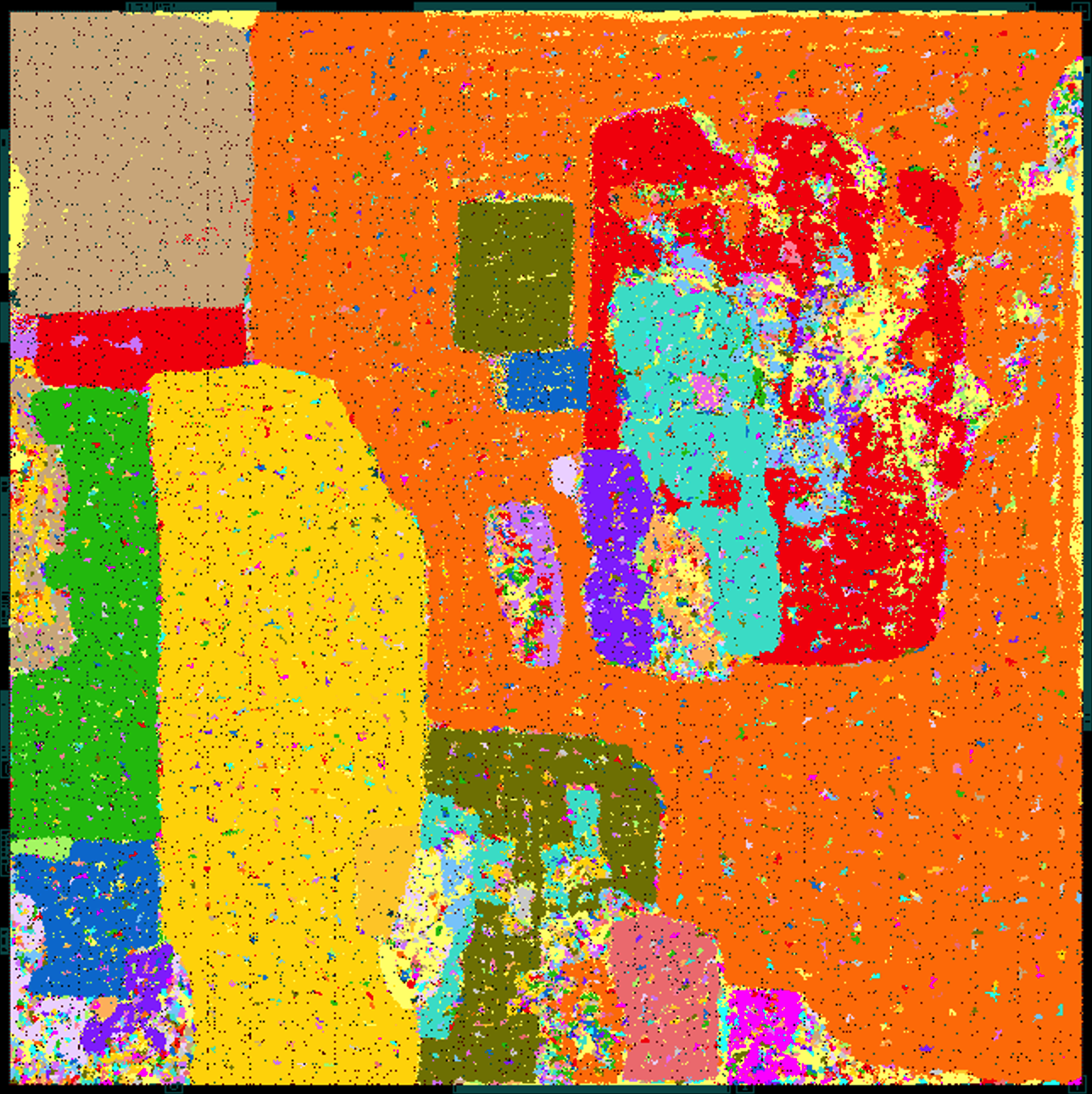}%
    }
    \caption{Layout pictures for FP and IB polar code decoders}
    \label{fig:layouts}
    
\end{figure}

Comparing the implementation results of the \IB decoder and the \SI{5}{\bit}
\FP decoder (Table~\ref{tab:1024}), the total area is reduced by \til \pc{15},
mostly stemming from the reduction in registers. This leads to an improved area
efficiency of \til\pc{18}, whereas the energy efficiency improves by
\til\pc{15}. Figure \ref{fig:layouts} shows the layouts of the \SI{5}{\bit} \FP
and the \SI{4}{\bit} \IB decoder.
It is also worth noting that, when compared to the close-to-float \SI{6}{\bit}
\FP decoder, accepting a small loss of \til \SI{0.2}{\decibel} in the error
correction leads to improvements of \til \pc{41} in area efficiency and
\til\pc{31} energy efficiency.

%% file: tables/results_128.tex
\begin{tabularx}{\textwidth}{l *{4}{C} }
\toprule
 &IB 4bit& \cite{giard2023}* & FP 4bit&FP 5bit \\
\midrule
& Place\&Route & Synthesis & Place\&Route & Place\&Route \\
Technology & \nm{12} & \nm{28}$\rightarrow$\nm{12} & \nm{12} & \nm{12} \\
\midrule
Frequency {[}MHz{]}       &500   & 1000 (3681)  &500 &500 \\
Throughput {[}Gbps{]}     &64    & 13 (47)      &64  &64  \\
Latency {[}ns{]}          &18    & 86 (23)          &18  &18  \\
Latency {[}CC{]}          &9     & 86          &9 &9 \\
Area {[}mm$^2${]}         &0.014 & 0.026        &0.012 &0.015 \\
- Registers               &0.005 & ---          &0.005 &0.006 \\
- Combinatorial           &0.005 & ---          &0.003 &0.005 \\
\textbf{Area Eff. {[}Gbps/mm$^2${]}}&\textbf{4528 }& \textbf{502 (1847)}  &\textbf{5384 }&\textbf{4248 }\\
Power Total {[}mW{]}      &21 & ---          &17 &24 \\
- Clock                   &7  & ---          &7  &9 \\
- Registers               &3  & ---          &3  &4 \\
- Combinatorial           &10 & ---          &7  &11 \\
\textbf{Energy Eff. {[}pJ/bit{]}}& \textbf{0.32 }& ---  &\textbf{0.27 }&\textbf{0.37 }\\
Power Density {[}W/mm$^2${]}&1.46 & --- &1.47 &1.58 \\
\bottomrule
\end{tabularx}

%% file: tables/results_1024.tex
\begin{tabular}{lcccc}
\toprule
 &IB 4bit&FP 4bit&FP 5bit&FP 6bit \\
\midrule
Frequency {[}MHz{]}       &750 &750 &750 &750 \\
Throughput {[}Gbps{]}     &768 &768 &768 &768 \\
Latency {[}CC{]}          &92 &92 &92 &92 \\
Area {[}mm$^2${]}         &0.822 &0.785 &0.968 &1.158 \\
- Registers               &0.360 &0.360 &0.439 &0.517 \\
- Combinatorial           &0.200 &0.168 &0.208 &0.274 \\
\textbf{Area Eff. {[}Gbps/mm$^2${]}}&\textbf{935 }&\textbf{979 }&\textbf{794 }&\textbf{663 }\\
Power Total {[}W{]}       &0.984 &0.866 &1.149 &1.431 \\
- Clock                   &0.208 &0.207 &0.254 &0.298 \\
- Registers               &0.280 &0.263 &0.354 &0.404 \\
- Combinatorial           &0.479 &0.381 &0.523 &0.706 \\
\textbf{Energy Eff. {[}pJ/bit{]}}&\textbf{1.28 }&\textbf{1.13 }&\textbf{1.50 }&\textbf{1.86 }\\
Power Density {[}W/mm$^2${]}&1.20 &1.10 &1.19 &1.24 \\
\bottomrule
\end{tabular}

%% file: 5_conclusion.tex
\section{Conclusion}
\label{sec:conclusion}

We presented fully characterized Fast-SSC Polar Decoders with an optimized \gls{IB}-based
quantization scheme. Especially for ultra-high throughput we outperform
decoders with comparable bit-width by \pc{18} in area efficiency and \pc{15} in
energy efficiency. This effect can be explained by the savings in memory
requirements of fully pipelined and unrolled decoders which is minimized with
the \gls{IB}-based quantization.